\documentclass[12pt]{iopart}
\usepackage{graphicx}
\expandafter\let\csname equation*\endcsname\relax
\expandafter\let\csname endequation*\endcsname\relax
\usepackage{amsmath}
\usepackage{color}
\graphicspath{{}}
\def\bra#1{\left\langle#1\right|}
\def\ket#1{\left|#1\right\rangle}

\begin{document}

\title[Three-dimensional Doppler, polarization-gradient, and magneto-optical forces]{Three-dimensional Doppler, polarization-gradient, and magneto-optical forces for atoms and molecules with dark states}

\author{J. A. Devlin and M. R. Tarbutt}
\address{Centre for Cold Matter, Blackett Laboratory, Imperial College London, Prince Consort Road, London SW7 2AZ, UK}

\begin{abstract}
We theoretically investigate the damping and trapping forces in a three-dimensional magneto-optical trap (MOT), by numerically solving the optical Bloch equations. We focus on the case where there are dark states because the atom is driven on a ``type-II'' system where the angular momentum of the excited state, $F'$, is less than or equal to that of the ground state, $F$. For these systems we find that the force in a three-dimensional light field has very different behaviour to its one dimensional counterpart. This differs from the more commonly used ``type-I'' systems ($F'=F+1$) where the 1D and 3D behaviours are similar. Unlike type-I systems where, for red-detuned light, both Doppler and sub-Doppler forces damp the atomic motion towards zero velocity,  in type-II systems in 3D, the Doppler force and polarization gradient force have opposite signs. As a result, the atom is driven towards a non-zero equilibrium velocity, $v_{0}$, where the two forces cancel. We find that $v_{0}^{2}$ scales linearly with the intensity of the light and is fairly insensitive to the detuning from resonance.  We also discover a new magneto-optical force that alters the normal MOT force at low magnetic fields and whose influence is greatest in the type-II systems. We discuss the implications of these findings for the laser cooling and magneto-optical trapping of molecules where type-II transitions are unavoidable in realising closed optical cycling transitions.
\end{abstract}

\maketitle

\section{Introduction}
Laser cooling \cite{Hansch1975} and magneto-optical trapping~\cite{Raab1987} are the foundations of a huge number of experiments and technological applications that use ultracold atoms. As an atom moves through the light field formed by counter-propagating red-detuned beams, its motion is damped due to the velocity-dependence of the Doppler shift. This is Doppler cooling, and it is effective in both an optical molasses and a magneto-optical trap (MOT). If the atom has angular momentum in the ground state, and the light field has a spatially varying polarization, the friction coefficient is modified at low velocities due to an interplay between optical pumping amongst the magnetic sub-levels and the changing polarization of the light~\cite{Dalibard1989, Ungar1989}. These sub-Doppler cooling mechanisms are usually divided into two types, depending on how the polarizations of the counter-propagating lasers are arranged. In the standard $\pi_{x}\pi_{y}$ configuration (also known as lin$\perp$lin), sub-Doppler cooling is due to the Sisyphus mechanism where a moving atom is optically pumped between magnetic sub-levels in such a way that it mostly climbs the potential hills arising from the spatially-varying light shift.  In the $\sigma^{+}\sigma^{-}$ configuration sub-Doppler cooling is due to an orientation-dependent force, where optical pumping induces an atomic orientation proportional to the speed of the atom, and the oriented atom absorbs more strongly from one beam than the other. Both sub-Doppler processes are inhibited by magnetic fields and so are most effective in an optical molasses where the field is close to zero, though they can also play a role in MOTs~\cite{Kohns1993a, Wallace1994, Townsend1995}. Sub-Doppler cooling can also occur without polarization gradients if the light has a spatially varying intensity and a suitable magnetic field is applied~\cite{Ungar1989, Sheehy1990, Shang1990, Gupta1994}. This is known as magnetically-induced laser cooling and can be viewed as a Sisyphus mechanism where the moving atom is transferred back and forth between magnetic sub-levels with differing light shifts, first by optical pumping in a region of high light intensity, and then by Larmor precession once it has moved towards low light intensity.

Most laser cooling work uses a ``type-I'' level system, where the atom is driven from a lower state with angular momentum $F$ to an upper state with angular momentum $F'=F+1$. This avoids optical pumping into dark states. The vast majority of MOTs use this level scheme and are known as type-I MOTs. Sub-Doppler cooling in type-I systems has been extensively studied experimentally~\cite{Weiss1989, Lett1989}, and theoretically, both in 1D~\cite{Dalibard1989, Ungar1989} and in 3D~\cite{Mo/lmer1991, Molmer1991, Javanainen1991, Javanainen1992, Javanainen1994}.

It is also possible to cool and trap atoms using a ``type-II'' transition where $F\geq F'$. In this case, atoms can be optically pumped into a dark state and uncouple from the light, and so these dark states must be destabilized~\cite{Berkeland2002} if the cooling is to be effective. Type-II MOTs have been studied experimentally~\cite{Raab1987, Prentiss1988, Shang1994, Flemming1997, Tiwari2008, Nasyrov2001, Atutov2001}. Compared to type-I MOTs, they tend to have higher temperatures, typically in the 2-20 mK range, larger cloud sizes, typically a few millimetres, and lower densities. These properties make them the poor relation of type-I atomic MOTs. However, type-II systems are increasingly being used to cool atoms efficiently to sub-Doppler temperatures in optical molasses, known as ``gray molasses''~\cite{Valentin1992, Boiron1995, Hemmerich1995, Grier2013, Burchianti2014, Sievers2015, Fernandes2012}.
Moreover, recent experiments on laser cooling~\cite{Shuman2010} and magneto-optical trapping \cite{Barry2014, McCarron2015, Norrgard2016} of molecules rely on type-II transitions to produce a closed optical cycling transition~\cite{Stuhl2008}. These advances have renewed the interest in the cooling and magneto-optical trapping mechanisms at work in type-II molasses and MOTs.

The theoretical understanding of the cooling and magneto-optical trapping forces in type-II systems is not as well developed as in the type-I case. Some insights into the behaviour of type-II MOTs in 3D are given in \cite{Tarbutt2015}, where rate equations are used to calculate the trapping and damping forces. This approach necessarily misses all of the sub-Doppler processes. There are a few theoretical studies of cooling in 1D that use a density matrix approach~\cite{Grier2013, Nienhuis1991, Grynberg1994, Weidemuller1994, Chang2002}. These studies point out that sub-Doppler processes occurring in type-II systems are different to those in type-I systems, because of the presence of dark states in the former. In a type-II system, the atom is optically pumped from a bright to a dark state, and then makes a motion-induced non-adiabatic transition back  to a bright state~\cite{Weidemuller1994, Fernandes2012}. This process, in combination with the spatially varying light shift can provide a frictional force. Some key conclusions of these 1D studies are: (i) Doppler cooling  always requires red-detuning, but sub-Doppler cooling in type-II systems requires blue-detuned light; (ii) there is no velocity-dependent force at any velocity in an $F=1 \rightarrow F'=1$ system in $\sigma^+\sigma^-$ light; (iii) the velocity-dependent forces tend to be considerably smaller in type-II systems compared to type-I. We know of no equivalent studies in 3D, with the exception of reference \cite{Sievers2015} where 3D simulations of gray molasses cooling on the D$_{1}$ line of Li and K are compared to experimental results.

Here, we systematically investigate the cooling and magneto-optical trapping forces for a number of type-II systems, namely $F\rightarrow F'$ = $1\rightarrow 0$,  $1\rightarrow 1$ and $2\rightarrow 1$. We also include $1\rightarrow 2$ transitions, for comparison between type-I and type-II systems. We numerically solve the optical Bloch equations, introduced in Section \ref{OBEs}, for an atom or molecule moving through either a 1D or 3D configuration of counter-propagating laser beams. In Section \ref{Cooling} we consider the cooling forces in zero magnetic field, and then extend this to consider cooling in the presence of a static magnetic field. Finally, in Section \ref{Trapping} we consider the trapping force in a MOT.

\section{Optical Bloch equations}
\label{OBEs}
\subsection{Equations of motion}

We follow the approach of Ungar et al. \cite{Ungar1989} in extending to multi-level atoms the theory first developed by Gordon and Ashkin \cite{Gordon1980}. We consider an atom with a ground-state of angular momentum $F$ and an excited state of angular momentum $F'$, separated by an energy $\hbar\omega_0$. The atomic states  $\ket{F,m_F}$ are labelled by the $F$ and $m_F$ quantum numbers. The equations of motion are set up in the Heisenberg picture, so we wish to find the time evolution of the expectation values of the atomic operators, $\left<\sigma^{i,j}_{m,n}\right>$. The operators for the ground and excited states are $\sigma^{F,F}_{m,n} = \ket{F,m} \bra{F,n}$ and $\sigma^{F',F'}_{m,n} = \ket{F',m} \bra{F',n}$ respectively. The operators for the ground-excited coherences rotate at $\omega$, the angular frequency of the laser light, and so we define $\sigma^{F,F'}_{m,n} = \ket{F,m} \bra{F',n}e^{i\omega t}$.  The Hamiltonian is
\begin{align}
\hat{H}=\hat{H}_\textrm{field}+\hat{H}_\textrm{atom}-\boldsymbol{\hat{\mu}}\cdot  \boldsymbol{\hat{B}} -\boldsymbol{\hat{d}}\cdot\boldsymbol{\hat{E}}\,\, ,
\end{align}
where
\begin{align}
\hat{H}_\textrm{field}=\frac{1}{2}\int \left(\varepsilon_0\boldsymbol{\hat{E}}^2+\frac{\boldsymbol{\hat{B}}^2}{\mu_0}\right)\, \text{d}V\, \, ,
\end{align}
and
\begin{align}
\hat{H}_\textrm{atom}=\frac{P^2}{2M}+\sum_m\hbar\omega_0\sigma^{F'F'}_{mm}\, \, .
\end{align}
\noindent Here $\boldsymbol{\hat{E}}$ is the electric field operator, $\boldsymbol{\hat{B}}$ is the magnetic field operator, $M$ is the atomic mass, $P$ is the momentum, $\boldsymbol{\hat{\mu}}$ is the magnetic dipole moment operator and $\boldsymbol{\hat{d}}$ is the electric dipole moment operator. Following the procedure in \cite{Ungar1989}, we expand the fields in terms of raising and lowering operators, and then write them in terms of the external field and the radiation reaction field of the atom. This leads to the following optical Bloch equations (OBEs):
\begin{align}
\frac{\text{d}\left<\sigma_{m, n}^{F, F}\right>}{\text{d} t}=& \sum_q\left[-(\Omega_{n}^q)^*\left<\sigma_{m, n+q}^{F, F'}\right>-\Omega_{m}^q\left<\sigma_{m+q,n}^{F', F}\right>+\Gamma^q_{m, n}\left<\sigma_{m+q, n+q}^{F', F'}\right>+\mathcal{B}^{F,F}_{m,n}\right]\label{eq:OBE1} ,\\
\frac{\text{d}\left<\sigma_{m, n}^{F', F'}\right>}{\text{d} t}=& \sum_q\left[(\Omega_{m-q}^q)^*\left<\sigma_{m-q, n}^{F, F'}\right>+\Omega_{n-q}^q\left<\sigma_{m,n-q}^{F', F}\right>-\Gamma\left<\sigma_{m, n}^{F', F'}\right>+\mathcal{B}^{F',F'}_{m,n}\right] ,\label{eq:OBE2}\\
\frac{\text{d}\left<\sigma_{m, n}^{F, F'}\right>}{\text{d} t}=& \sum_q\left[\Omega_{n-q}^q\left<\sigma_{m, n-q}^{F, F}\right>-\Omega_{m}^q\left<\sigma_{m+q,n}^{F', F'}\right>+\left(i\Delta-\Gamma/2\right)\left<\sigma_{m, n}^{F, F'}\right>+\mathcal{B}^{F,F'}_{m,n}\label{eq:OBE3}\right].
\end{align}
The detuning is $\Delta=\omega-\omega_0$, the total decay rate from the excited state is $\Gamma$, and the relaxation rates $\Gamma^q_{m, n}$ are given by
\begin{align}
\Gamma^q_{m, n}=&\Gamma(2F'+1) (-1)^{-m-m'}\begin{pmatrix}
 F & 1 & F' \\
    -m & -q & m+q
\end{pmatrix}\begin{pmatrix}
 F & 1 & F' \\
    -n & -q & n+q
\end{pmatrix}\, .
\end{align}
The Rabi frequencies $\Omega^{q}_{n}$ are
\begin{align}
\Omega^{q}_{n}=&\frac{i}{2\hbar} \mathcal{E}_{-q}(\boldsymbol{r})(-1)^{F-n}\begin{pmatrix}
 F & 1 & F' \\
    -n & -q & n+q
\end{pmatrix}\langle F'\|d\|F\rangle\, , \label{eq:rabirate}
\end{align}
where $\langle F'\|d\|F\rangle$ is the reduced matrix element of the electric dipole operator. Here, $\mathcal{E}^{}_q(\boldsymbol{r})$ are amplitude components of the total classical electric field $\boldsymbol{E}(\boldsymbol{r},t)=\boldsymbol{\mathcal{E}}(\boldsymbol{r})\cos\omega t$, expanded in a spherical basis according to ${\boldsymbol{\mathcal{E}}(\boldsymbol{r})=\sum_q\mathcal{E}^{}_q(\boldsymbol{r})\epsilon^*_q}$, and $\textbf{r}$ is the position of the atom. The spherical basis vectors are $\epsilon^{}_0=\textbf{e}_z$, $\epsilon^{}_{\pm1}=\mp(\textbf{e}_x\pm i \textbf{e}_y)/\sqrt{2}$.
The effect of the Zeeman interaction is contained in the expression
$\mathcal{B}^{i,j}_{m,n}$, which is
\begin{align}
\mathcal{B}^{i,j}_{m,n}=\frac{1}{2}\sum_q\bigl[&-B_q(\textbf{r})\left((\mu_{n-q}^{q,j})^*\left<\sigma_{m, n-q}^{i, j}\right>-(\mu_{m}^{q,i})^*\left<\sigma_{m+q, n}^{i, j}\right>\right)\nonumber \\
&\,\,+ B^*_q(\textbf{r})\left(\mu_{n}^{q,j}\left<\sigma_{m, n+q}^{i, j}\right>-\mu_{m-q}^{q,i}\left<\sigma_{m-q, n}^{i, j}\right>\right)\bigr] \, ,
\end{align}
where $B^{}_q(\boldsymbol{r})$ are the spherical components of the magnetic field amplitude and the magnetic moment terms are
\begin{align}
\mu_{n}^{q,j}=-\frac{i}{\hbar}g(j)\mu_\textrm{B}(-1)^{j-n}(-1)^q\sqrt{j(j+1)(2j+1)} \begin{pmatrix}
    j & 1 & j \\
    -n & -q & n+q
  \end{pmatrix}\, \, ,
\end{align}
with $g(F)=g$ and $g(F')=g'$, the lower and upper state g factors respectively.

The force $\textbf{f}=\text{d} \textbf{P}/\text{d} t$ is treated classically. This is justified as long as the temperature remains high compared to the photon recoil limit $T_\textrm{recoil}=(h/\lambda)^2/(2Mk_B)$ where the effects of momentum quantisation become significant. The force can be expressed as a commutator of the momentum $\textbf{P}$ with the Hamiltonian, which can also be re-written for this Hamiltonian as $\textbf{f}=-\nabla \hat{H}$.
The direct contribution from gradients in the magnetic field interaction
$\nabla (\boldsymbol{\hat{\mu}}\cdot\textbf{B})$ is small for typical MOT fields, so we neglect it here, but emphasise that the magnetic field can give rise to substantial forces indirectly, by creating unbalanced radiation pressure. The force arising from the electric dipole interaction is
\begin{align}
\textbf{f}&=-i\hbar\sum_{m,q}\left[\left<\sigma_{m+q,m}^{F',F}\right>\nabla\Omega_m^q-\left<\sigma_{m,m+q}^{F,F'}\right>(\nabla\Omega_m^q)^*\right].\label{eq:force}
\end{align}

\subsection{Light fields}
In one dimension, we will focus on two light fields that give interesting polarization gradient forces, the  lin$\phi$lin and $\sigma^+\sigma^-$ configurations. The 1D lin$\phi$lin configuration consists of two counter-propagating linearly polarised beams, with an angle $\phi$ between the polarisations. Both beams have amplitude $E$ and wavevector $k=\omega/c$, and the total light field is
\begin{align}
\boldsymbol{\mathcal{E}}_{\text{lin}\phi\text{lin}}(z)=E\sqrt{2}\left[\cos(kz-\phi/2)\epsilon^{}_{-1}-\cos(kz+\phi/2)\epsilon^{}_{1}\right]\, . \label{eq:firstStandingWave}
\end{align}
We will also consider a 3D version of the standing wave where additional pairs of counter-propagating linear beams propagate along the $x$ and $y$ axes and $\phi=\pi/2$ for each pair of beams. The resulting light field is
\begin{align}
\boldsymbol{\mathcal{E}}_{\text{lin}\perp\text{lin}3\text{D}}(\boldsymbol{r})=E\left[(\textbf{e}_xe^{i kz}+\textbf{e}_ye^{-i kz})+e^{i \theta_1}(\textbf{e}_ze^{i ky}+\textbf{e}_xe^{-i ky})+e^{i \theta_2}(\textbf{e}_ye^{i kx}+\textbf{e}_ze^{-i kx})\right]\, . \label{eq:linperplin3D}
\end{align}
The phases $\theta_1$ and $\theta_2$ arise because each additional pair of counter-propagating beams can have an arbitrary phase with respect to the beams which propagate along $z$ \cite{Hopkins1997}.  

The 1D $\sigma^+\sigma^-$ configuration consists of two counter-propagating beams with identical circular polarization (relative to their $k$-vectors),
\begin{align}
\boldsymbol{\mathcal{E}}_{\sigma^+\sigma^-}(z)=E \sqrt{2}\left[\textbf{e}_x\sin(kz)-\textbf{e}_y\cos(kz)\right]\, .
\end{align}
In the same way as we defined equation~(\ref{eq:linperplin3D}) we can also make a 3D version of the $\sigma^+\sigma^-$ arrangement
\begin{align}
\boldsymbol{\mathcal{E}}_{\sigma^+\sigma^-3\text{D}}(\boldsymbol{r})=E \sqrt{2}&\left[(\textbf{e}_x\sin(kz)-\textbf{e}_y\cos(kz))+e^{i \theta_1}(\textbf{e}_y\sin(kx)-\textbf{e}_z\cos(kx))\right. \nonumber\\
&\,\,\left.+e^{i \theta_2}(\textbf{e}_z\sin(ky)-\textbf{e}_x\cos(ky))\right]\, .
\label{eq:spsm3dStandingWave}
\end{align}

The intensity of each laser beam is $I=\tfrac{1}{2}c\varepsilon_0E^2$, so the total intensity is $2I$ for the 1D arrangements and $6I$ for the 3D arrangements. It is helpful to specify the intensity relative to the saturation intensity of the transition, defined via
\begin{align}
\frac{I}{I_{\rm{sat}}}=\frac{2E^2|\langle F'\|d\|F\rangle|^2}{\hbar^2\Gamma^2(2F'+1)}\, . \label{eq:lastStandingWave}
\end{align}
This definition leads to the usual, convenient expression for the saturation intensity, ${I_{\rm{sat}}=\frac{\pi h c \Gamma}{3\lambda^3}}$.

\section{Cooling forces}
\label{Cooling}
\subsection{1D cooling in zero magnetic field}
To investigate the cooling force, it is natural to start with the simple 1D light fields $\boldsymbol{\mathcal{E}}_{\sigma^+\sigma^-}$ and $\boldsymbol{\mathcal{E}}_{\text{lin}\phi\text{lin}}$. We drag the atom through the light field at a velocity $v=z/t$. In the $\sigma^+\sigma^-$ case we wait for the expectation values of the atomic operators to reach a steady state and then calculate the $z$ component of the force. In the lin$\phi$lin case, we wait until the expectation values reach a periodic quasi-steady state, and then calculate the $z$ component of the force averaged over one oscillation period of the atomic operators. The force parallel to the velocity is positive if it is in the same direction as the velocity vector. With this terminology, a normal friction force is a negative parallel force. Throughout this paper, we express the velocity in units of $\Gamma/k$ and the force in units of $\hbar k \Gamma$. For Rb cooled on the D$_2$ transition, the Doppler-limited temperature, $T_\textrm{D}=\hbar \Gamma/(2 k_B)$, corresponds to an rms speed of $0.04\Gamma/k$.

\begin{figure}[htb]
\includegraphics[scale=1]{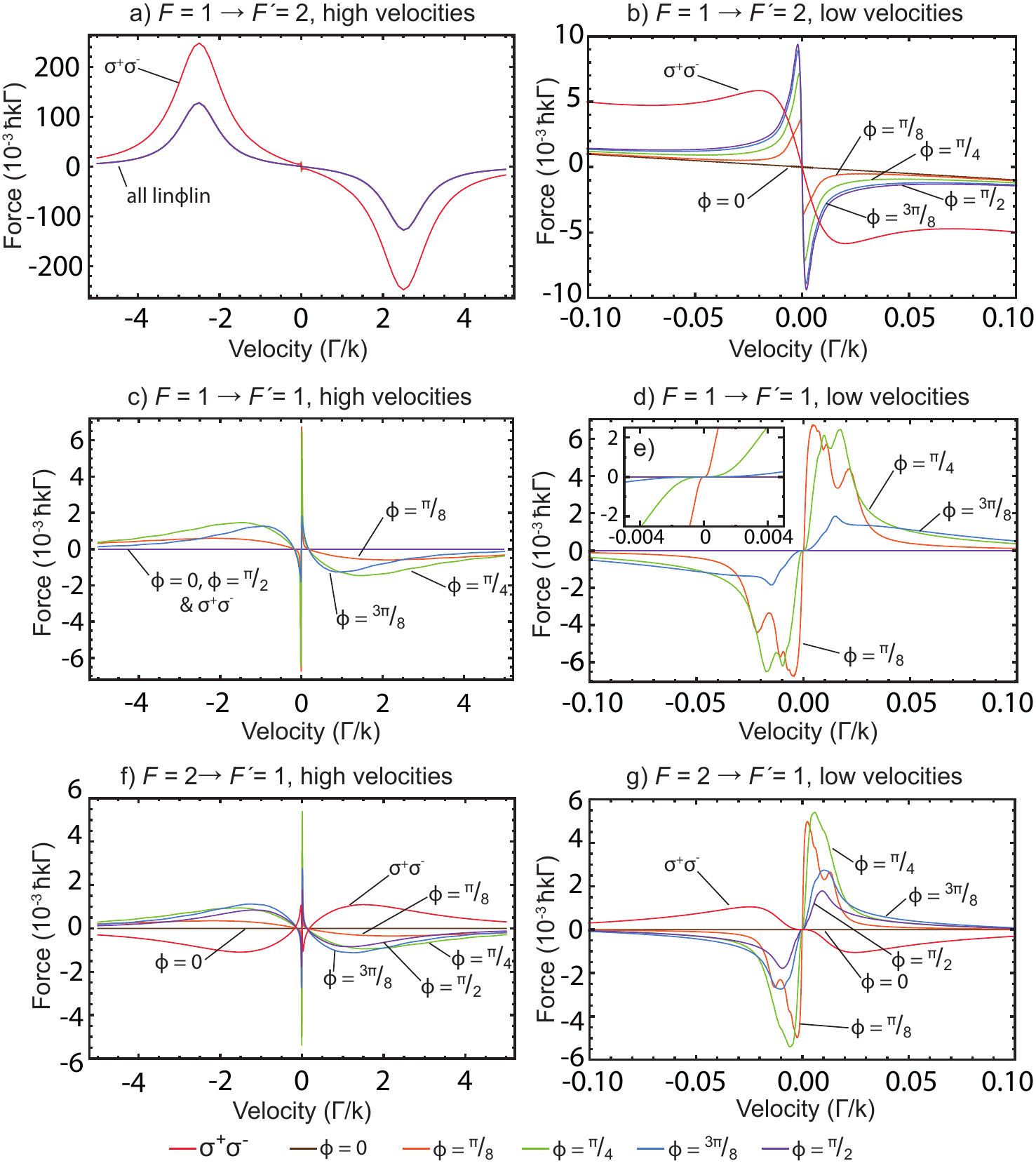}
\caption{(Quasi)-steady state force versus velocity for the 1D $\sigma^+\sigma^-$ arrangement and lin$\phi$lin arrangements for atoms scattering light on  $F=1\rightarrow F'=2$, $F=1\rightarrow F'=1$ and   $F=2\rightarrow F'=1$ transitions. The parameters are $\Delta=-2.5\Gamma$ and $I=I_{\rm{sat}}$ for each beam.}
\label{fig:1DResults}
\end{figure}

Figures~\ref{fig:1DResults}~a) and b) show the force versus velocity for the $1\rightarrow 2$ type-I system in both the $\sigma^+\sigma^-$ light field and the $\text{lin}\phi\text{lin}$ configuration for several values of $\phi$. For all polarization configurations this system exhibits the usual Doppler cooling curve, with maximum force when the Doppler shift equals the detuning (here, $\Delta=-2.5\Gamma$). The sub-Doppler forces are shown most clearly in figure~\ref{fig:1DResults}~b). When $\phi=0$ there is no sub-Doppler force because there are no polarization gradients, but for $\phi\neq0$ and for the $\sigma^+\sigma^-$ light field there is enhanced friction at low velocities. The highest peak sub-Doppler force occurs when the beams are orthogonally polarised, i.e. $\phi=\pi/2$. Relative to this case, the $\sigma^+\sigma^-$ case shows larger Doppler and smaller sub-Doppler features.

The story is very different for the type-II systems. Figures~\ref{fig:1DResults}~c), d) and e) show the force versus velocity for the $1\rightarrow 1$ system for the same light fields as figure~\ref{fig:1DResults}~a). In this system the force is zero at all velocities for the $\sigma^+\sigma^-$ configuration, and for the $\text{lin}\phi\text{lin}$ configuration with $\phi=0$ and $\phi=\pi/2$, in agreement with the results described in \cite{Weidemuller1994, Chang2002, Cohen-Tannoudji1992, Nasyrov2001}. For other values of $\phi$, there is a force at low velocity of approximately the same size as the sub-Doppler force in the type-I $1\rightarrow 2$ system, but of opposite sign. Rather than depending linearly on the velocity around $v=0$, this force scales with a higher odd power of $v$, as can be seen clearly in figure~\ref{fig:1DResults}~e). There is also a broad frictional feature at velocities of around $\Gamma/k$ which is roughly 100 times smaller than the Doppler cooling feature in the $1\rightarrow 2$ system. These results show that, for the $1\rightarrow 1$ system, there can be strong sub-Doppler cooling at low velocity for blue-detuned light, but only weak cooling at higher velocities for red-detuned light.

The marked difference in the velocity-dependent force between the $1\rightarrow 2$ and $1\rightarrow 1$ systems can be attributed to the presence of a dark state in the $1\rightarrow 1$ system. At high velocities of order $v\sim\Delta/k$  the dark state prevents normal Doppler cooling in the $1\rightarrow 1$ system. At these high velocities, the Doppler shift is large enough that the beams can be considered independently. One of the two beams optically pumps the atom into a state which is dark to that beam. If $\phi \ne 0$ the opposing beam can return the atom to the state which is bright to the first beam.  For these two processes --- optical pumping into the dark state by one beam and out of the dark state by the opposing beam --- the average number of scattered photons is the same, and so the net force is zero. This explains the absence of a Doppler cooling feature in figure~\ref{fig:1DResults}~c).

\begin{figure}[tb]
\includegraphics[scale=1]{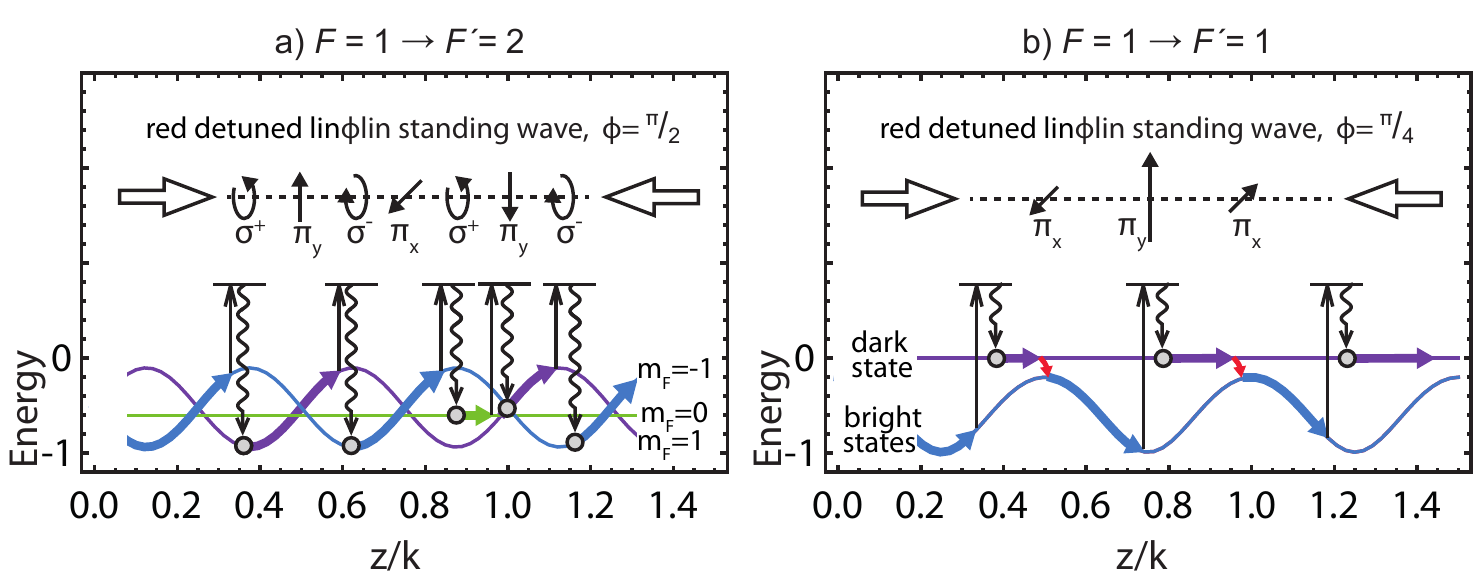}
\caption{Illustration of the Sisyphus mechanism for a red-detuned lin$\phi$lin standing wave in two cases: a) $1\rightarrow2$ system with $\phi=\pi/2$ and b) $1\rightarrow1$ system with $\phi=\pi/4$. The upper parts show the local polarisation and the lower parts the ac Stark shifts of the ground states. In a) the mechanism relies only on optical pumping between differently-shifted sub-levels. In b) the mechanism relies on optical pumping near the intensity maxima and non-adiabatic transitions from the dark to the bright state near the intensity minima.}
\label{fig:1DExplantion}
\end{figure}

At low velocities where $v\ll \Gamma/k$, the dark state has a different impact, which is best illustrated using a picture of how the polarization gradient force arises in a red-detuned $\text{lin}\phi\text{lin}$ standing wave, for both the $1\rightarrow 2$ and $1\rightarrow 1$ systems. Figure~\ref{fig:1DExplantion}~a) illustrates the familiar Sisyphus cooling mechanism \cite{Dalibard1989} for the $1\rightarrow 2$ system in the lin$\perp$lin configuration (i.e. $\phi=\pi/2$, the angle which maximizes the force). The bottom part of the figure show the ac Stark shifts of the $m_F=0,\pm1$ ground states, along with a sequence of optical pumping events. The thick lines indicate the internal state of an atom as it moves through the standing wave, showing that the atom loses energy because in a red-detuned light field the state with the largest negative ac Stark shift (and hence the lowest energy) is also the state into which the atom is continually being optically pumped.

The picture is different for the $1\rightarrow1$ system in a lin$\phi$lin configuration, as pointed out previously \cite{Weidemuller1994}  \cite{Shahriar1993}. Here, the mechanism is best explained in terms of a position-dependent dark state and an orthogonal bright state. Figure~\ref{fig:1DExplantion}~b) shows the ac stark shifts of these states as a function of position, for the case where $\phi=\pi/4$, along with the direction of the local laser polarisation at various points in the standing wave. An atom that starts in the bright state will be pumped into the dark state, and this is most likely to occur at the intensity maxima of the standing wave. If the atom is stationary it will remain in the dark state, and there will be no force, but if it is moving through the changing polarization it may make a non-adiabatic transition back into the bright state. The probability of such a transition is strongly peaked at places where the energy difference between bright and dark states is small, and this occurs at the intensity minima of the standing wave where the energy of the bright state is a maximum. This picture shows how the atom gains energy by repeatedly transitioning to the bright state at the top of the potential hill, and then being optically pumped to the dark state at the bottom of the hill. When the light is blue-detuned the ac Stark shift is reflected through the zero energy line, and the force instead becomes a frictional one that cools the atom. We note that this ``non-adiabatic force''~\cite{Weidemuller1994} is zero when $\phi=0$ and when $\phi=\pi/2$. When $\phi=0$ there are no polarization gradients and so the non-adiabatic transition probability is everywhere zero. When $\phi=\pi/2$ both the ac Stark shift and the non-adiabatic transition probability are independent of position. Intermediate values of $\phi$ between 0 and $\pi/2$ show a difference in the size of the non-adiabatic force and the velocity range over which it acts. When $\phi$ is small the friction force is strong at low velocities but acts over a small velocity range, while the opposite is true for values of $\phi$ close to $\pi/2$. A smaller value of $\phi$ increases the non-adiabatic transition probability, and this increases the force for the slowest molecules which only make those transitions near the nodes. For the same reason, faster molecules can make the transition away from a node, and may then ride over the top of the next potential hill before being optically pumped back to the dark state, resulting in a decreased friction force. The $\phi$ that maximizes the sub-Doppler force is between $\pi/8$ and $3\pi/16$ for the $F=1 \rightarrow F'=1$ system, and between $3\pi/16$ and $\pi/4$ for the $F=2 \rightarrow F'=1$ system.

We have also studied the $1\rightarrow 0$ and $2\rightarrow 1$ systems in these one-dimensional light fields. For the $1\rightarrow 0$ system the force is zero at all velocities in the $\sigma^+\sigma^-$ case and for all values of $\phi$ in the lin$\phi$lin case. The Doppler cooling force is zero for the same reason as in the $1\rightarrow 1$ case discussed above. The non-adiabatic force is zero because the motion through the changing polarization does not couple the bright and dark states (the rate of change of the dark state is orthogonal to the bright state). The results for the $2\rightarrow1$ system are shown in figure~\ref{fig:1DResults}~f) and g). Here, the force is non-zero in all cases except for the lin$\phi$lin configuration with $\phi=0$. In this system there are three bright states and two dark states, and while the ac Stark shift of one of the bright states is zero at $\phi=\pi/2$, the other two are position-dependent and so the non-adiabatic force acts. Interestingly, the force in the $\sigma^+\sigma^-$ configuration has a similar size to the lin$\phi$lin configuration, but has the opposite sign. The ac  Stark shifts of the bright states in the $\sigma^+\sigma^-$ case are independent of position, and so the non-zero force indicates that a cooling mechanism can operate in some type-II systems in $\sigma^+\sigma^-$ light fields that is similar to the orientation-dependent force that acts in type-I systems.

Although these 1D results are not new, we present them here for two reasons. First, they validate our code, because our results agree with those given in Ref.~\cite{Chang2002}. We have also confirmed that our results match those of Ref.~\cite{Ungar1989} for the $2\rightarrow3$ system. Second, the 1D results serve as a benchmark for comparison with the 3D results presented in the next section. We will find that the 1D results provide a good description of the equivalent 3D geometries for the type-I systems, but that type-II systems in 3D differ greatly from their 1D counterparts.

\subsection{3D cooling in zero magnetic field}

We now consider the velocity-dependent forces in 3D. Once again the atom is dragged through the light field, from an initial position $\boldsymbol{r}_0$ to a new position $\boldsymbol{r}=\boldsymbol{r}_0+\boldsymbol{v}t$ at time $t$. Provided that the components of $\boldsymbol{v}$, $(v_x, v_y, v_z)$, have a greatest common divisor $g$, the Hamiltonian will be periodic and the atomic operators and force will come to a quasi-steady state, where they too become periodic in time with a time period equal to $\lambda/g$. To extract the steady state force, we evolve the OBEs until the system has come into a quasi-steady state, then evaluate the atomic operators and components of the force over one period $\lambda/g$. The steady state force is the average of the force over one period and is found to be independent of the initial values of the atomic operators, which we take to be a maximally mixed state of the ground state operators. In 3D, we distinguish between components of the force that are parallel and perpendicular to the velocity vector.

In three dimensions, the intensity and polarization gradients in both $\sigma^+\sigma^-$ and lin$\perp$lin light fields depend on the relative phases $\theta_1$ and $\theta_2$ between the three pairs of counter-propagating beams. The electric field experienced by the atom depends on its starting position, $\boldsymbol{r}_0$, and its direction of motion $\boldsymbol{v}/|\boldsymbol{v}|$. Therefore, at low velocities we expect the force to depend on speed, position and direction. At higher velocities where the Doppler forces dominate, we no longer expect a significant dependence on position, but the direction of motion remains important. This is because the net Doppler shift depends on the orientation of the trajectory relative to the $k$-vectors of the light beams.

In a typical MOT or molasses, the atomic cloud extends over many thousands of spatial periods of the light field. To find the steady-state force at each velocity we solve the OBEs for a set of initial positions $\boldsymbol{r}_0$ and then average together the results. In our first approach to this problem, we choose a set of points evenly spaced by $\lambda/8$ along each Cartesian axis and filling a cube of volume $\lambda^{3}$, giving 512 points in total.

\begin{figure}[tb]
\begin{center}
\includegraphics[scale=1]{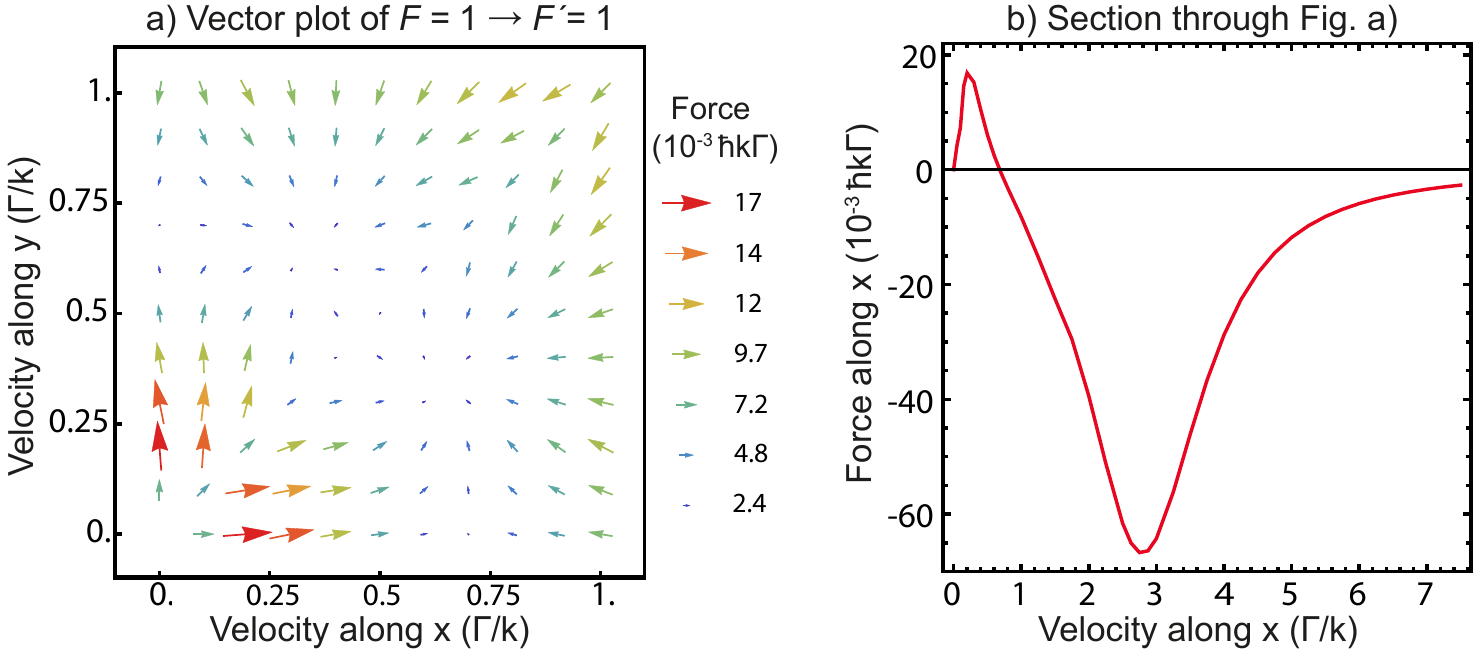}
\end{center}
\caption{Steady state force in the $xy$ plane for a $1\rightarrow 1$ system moving in a 3D $\sigma^+\sigma^-$ light field given by $\boldsymbol{\mathcal{E}}_{\sigma^+\sigma^-3\text{D}}(\boldsymbol{r})$. Each choice of $\boldsymbol{v}$ plotted has been averaged over 512 values of $\boldsymbol{r}_0$ as described in the text. The vector plot a) shows the $xy$ component of the force as a function of $v_x$ and $v_y$, while plot b) fixes $v_y=0$ and plots the $x$ component of the force as a function of $v_x$, considering higher speeds than those shown in plot a).  Other parameters are  $v_z=0.1\Gamma/k$, $\Delta=-2.5\Gamma$ and $I=2I_\textrm{sat}$ for each beam.}
\label{fig:3DResultsArrows}
\end{figure}

Figure~\ref{fig:3DResultsArrows} shows the outcome of this approach for a $1\rightarrow1$ system moving in a 3D $\sigma^+\sigma^-$ lattice with $\theta_1=\theta_2=0$. In figure~\ref{fig:3DResultsArrows}~a), we show a vector plot of the steady-state force in the $xy$ plane in the case where $v_z=0.1\Gamma/k$, while $v_x$ and $v_y$ are varied. The detuning of all beams is $\Delta=-2.5\Gamma$ and the intensity of each beam is $2I_\textrm{sat}$. Notice first that unlike the 1D $\sigma^+\sigma^-$ standing wave, in 3D the force does not vanish. The figure shows that at high velocities the arrows predominantly point towards the origin, meaning that a negative force dominates.  This is normal Doppler cooling, which can occur in 3D because the laser beams transverse to the motion can pump population out of the dark state into a bright state. At low velocities there is a positive force, tending to push the atoms to higher velocities rather than cool them. This is the influence of the polarization gradient force. There are several velocities where the force in the $xy$-plane is zero. These are (to the nearest 0.025) $(v_x,v_y) = (0,0)$, $(0,0.7)$, $(0.7,0)$, $(0.45,0.45)$, $(0.35,0.65)$ and $(0.65,0.35)$, all in units of $\Gamma/k$. The first four are unstable points in this plane, while the last two are stable  in this plane in the sense that a small deviation from the equilibrium  $xy$-velocity produces a force that returns the velocity to the equilibrium value. We see that for this system red-detuned light does not cool the atom towards zero velocity but instead drives it towards particular non-zero values. This is seen clearly in figure~\ref{fig:3DResultsArrows}~b) where the force along $x$ is plotted versus $v_x$, when $v_{y}=0$, and $v_{z}=0.1\Gamma/k$. Finally, returning to figure~\ref{fig:3DResultsArrows}~a) we note that at low velocities, the perpendicular part of the force becomes comparable to the parallel force, even after averaging over the initial positions. The non-vanishing perpendicular components are also seen in the type-I systems ($1\rightarrow2$ and $2\rightarrow3$) studied in \cite{Mo/lmer1991,Molmer1991}. These perpendicular components may play an important role in sub-Doppler cooling and the dynamics of a MOT, and could lead to non-isotropic velocity distributions. Nevertheless, in this paper we focus on the parallel component of the force and we assume an isotropic velocity distribution.

To calculate representative velocity-dependent force curves, similar to those in figure~\ref{fig:1DResults}, we adopt a Monte-Carlo approach. We pick a starting position $\boldsymbol{r}_0$ whose Cartesian components are drawn at random from a uniform distribution on the interval $[0,\lambda]$, and a random direction of motion chosen from an isotropic distribution. We then find the steady-state force parallel to the velocity $\boldsymbol{f}\cdot\boldsymbol{v}/|\boldsymbol{v}|$ for a range of speeds $|\boldsymbol{v}|$. We repeat this procedure many times, typically making 500-5000 choices of initial position and direction. We average together the results of the simulations at each speed, and determine the confidence intervals on the mean force using a bootstrap procedure \cite{BradEandTibshirani1993}. In this paper, we restrict ourselves to the case where $\theta_1=\theta_2=0$; alternative choices for the phases can produce slightly different force curves but the conclusions of the subsequent sections are unaffected.   

\begin{figure}[tb]
\begin{center}
\includegraphics[scale=1]{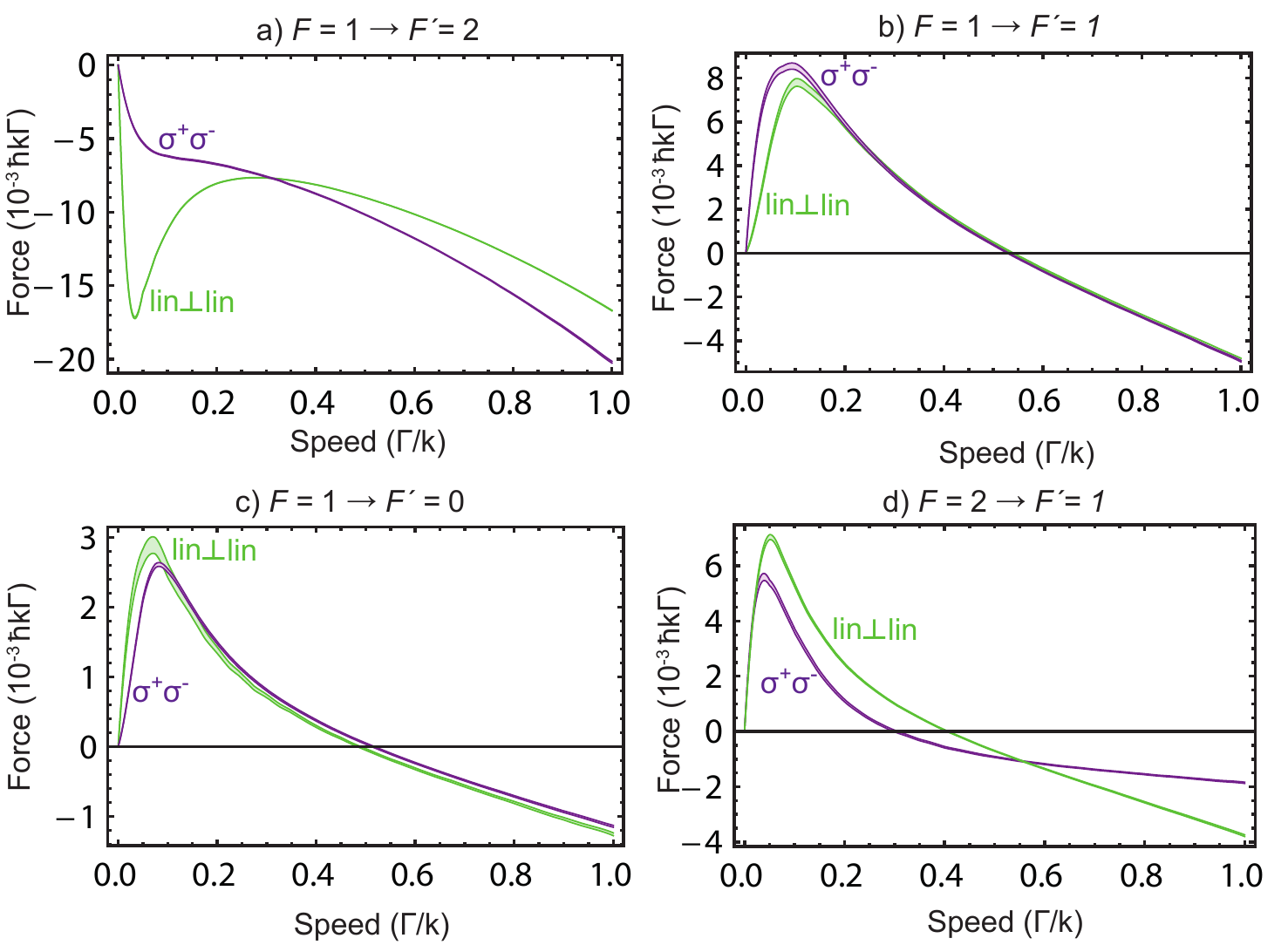}
\end{center}
\caption{Steady-state force in the direction of velocity for a) $1\rightarrow 2$; b) $1\rightarrow 1$; c) $1\rightarrow 0$; d) $2\rightarrow 1$ systems moving in a 3D $\sigma^+\sigma^-$ light field (purple) and 3D $\textrm{lin}\perp\textrm{lin}$ light field (green). Parameters are $\Delta=-2.5\Gamma$ and $I=I_\textrm{\rm{sat}}$ for each beam. The width of the lines indicates the $68\%$ confidence intervals on the mean force, calculated using a bootstrap method.}
\label{fig:ResultsIn3D}
\end{figure}

Figure~\ref{fig:ResultsIn3D} shows the results of such a simulation for the $1\rightarrow 2$, $1\rightarrow 1$,  $1\rightarrow 0$ and $2\rightarrow 1$ systems, for the 3D $\sigma^+\sigma^-$ and lin$\perp$lin arrangements. The light is detuned to the red ($\Delta=-2.5\Gamma$) and the intensity of each beam is $I_\textrm{sat}$. The $1\rightarrow2$ system shows an enhanced frictional force at low speeds due to sub-Doppler cooling, and this effect is more pronounced for lin$\perp$lin than $\sigma^+\sigma^-$. These 3D results are very similar indeed to their 1D counterparts. We turn now to the type-II systems shown in figure~\ref{fig:ResultsIn3D}~b)-d).  These results are completely different to the 1D results, and also to the type-I systems in 3D. For all three systems, and for both polarization configurations, the force in 3D is positive at low speeds and negative at speeds higher than around $\Gamma/2k$. This behaviour is reversed when the detuning is reversed. The forces are large, typically only slightly smaller than for type-I systems. This contrasts sharply with the 1D case where there is no velocity-dependent force for $1\rightarrow 0$ and $1\rightarrow 1$ systems in the 1D versions of these light fields, and only small forces for the $2\rightarrow1$ system. In the type-II systems there is not much difference between the $\sigma^+\sigma^-$ and lin$\perp$lin cases, unlike the type-I systems where the difference is pronounced at low speeds. Furthermore, the low velocity features have a significant influence on the force up to higher speeds in type-II systems compared to type-I systems, especially in the $\sigma^+\sigma^-$ configuration. In fact the ``sub-Doppler'' features we see in type-II systems extend far beyond the usual Doppler limit even for moderate saturation of the transition. This result suggests that, for red-detuned light and typical intensities, atoms and molecule cooled on type-II transitions will reach an equilibrium temperature much higher than the Doppler limit.

We attribute this difference between the 1D and 3D results to the non-adiabatic force, which is effective in 3D $\sigma^+\sigma^-$ and lin$\perp$lin fields but not in their corresponding 1D versions. This is because, on any typical trajectory through these 3D fields, there are gradients of both intensity and polarization, giving rise to spatially varying ac Stark shifts, optical pumping probabilities and non-adiabatic transition probabilities, all the ingredients needed for the non-adiabatic force. For blue-detuned light the sign of the force is reversed, and so the non-adiabatic force leads to sub-Doppler cooling. The role of the non-adiabatic force in sub-Doppler cooling of type-II systems was elucidated in Ref.~\cite{Fernandes2012}, in the context of an experimental demonstration of gray molasses cooling of $^{40}$K.

\begin{figure}[tb]
\begin{center}
\includegraphics[scale=1]{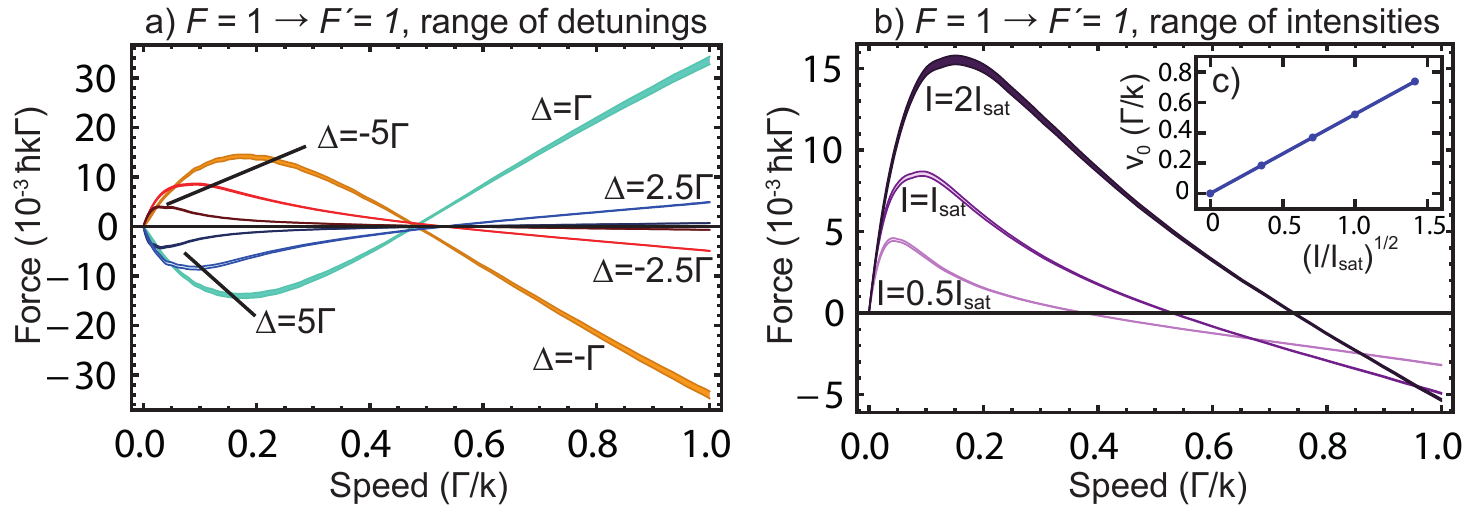}
\end{center}
\caption{Steady state force in the direction of velocity for a $1\rightarrow 1$ system moving in a 3D $\sigma^+\sigma^-$ optical standing wave for a) a range of detunings with an intensity of $I=I_\textrm{sat}$ in each beam and b) a range of intensities at a fixed detuning $\Delta=-2.5\Gamma$. Inset c) shows $v_0$, the speed where the force is zero, as a function of the square root of the intensity.}
\label{fig:DetuningResults3D}
\end{figure}

We now examine how the friction force depends on detuning and intensity. Figure~\ref{fig:DetuningResults3D}~a) shows the force versus velocity in the $1\rightarrow1$ system for a range of detunings when the intensity of each laser beam is $I=I_\textrm{sat}$. We see that reversing the sign of the detuning reverses the sign of the force in the usual way. Both the peak sub-Doppler force visible in figure~\ref{fig:DetuningResults3D}~a) and the peak Doppler force (which occurs at higher speeds than those shown in this figure) scale in the same way with detuning; for large detunings the forces scale as $\frac{1}{\Delta^2}$, reaching a maximum at around $\Delta=\Gamma$ and then falling as the detuning is further reduced. For small detunings in the range $0$--$\Gamma/2$, the peak forces scales linearly with detuning.   When $I=I_\textrm{sat}$, the force passes through zero at velocity $v_{0} \approx 0.5 \Gamma/k$. The velocity $v_0$ depends only weakly on the detuning, varying by only 15\% over the entire range from $\Delta=-10\Gamma$ to $\Delta = -0.125\Gamma$ that we considered. The behaviour is similar in other type-II systems. 
Figure~\ref{fig:DetuningResults3D}~b) shows the force versus velocity for a range of intensities with the detuning fixed at $\Delta=-2.5\Gamma$. Over the range of intensities shown, the peak value of the sub-Doppler force is proportional to the intensity, while the slope of the force close to zero speed is independent of intensity just as it is for type-I systems~\cite{Ungar1989}. The speed, $v_{0}$, where the force changes sign, scales with the square root of the intensity, as shown in figure~\ref{fig:DetuningResults3D}~c). In a red-detuned type-II MOT the balance between Doppler and sub-Doppler forces drives atoms towards speed $v_{0}$. Therefore, we can expect the temperature, which is proportional to the square of the speed, to scale linearly with intensity, until the intensity is low enough that the Doppler-limiting temperature is reached. This is the behaviour recently observed in a SrF MOT \cite{Norrgard2016}.

\subsection{3D cooling in a magnetic field}

It is well known that, in type-I systems, sub-Doppler cooling is suppressed when a magnetic field is applied~\cite{Walhout1992}, and so most sub-Doppler cooling is done in a molasses rather than a MOT. We wish to know whether this is also the case for type-II systems, and so we repeat the analysis of the previous section with the addition of a homogeneous magnetic field along the $z$ axis\footnote{The same behaviour is also seen for most other orientations of the magnetic field with respect to the laser beam propagation direction.}, $B_q=B\epsilon_0$.
\begin{figure}[htb]
\begin{center}
\includegraphics[scale=1]{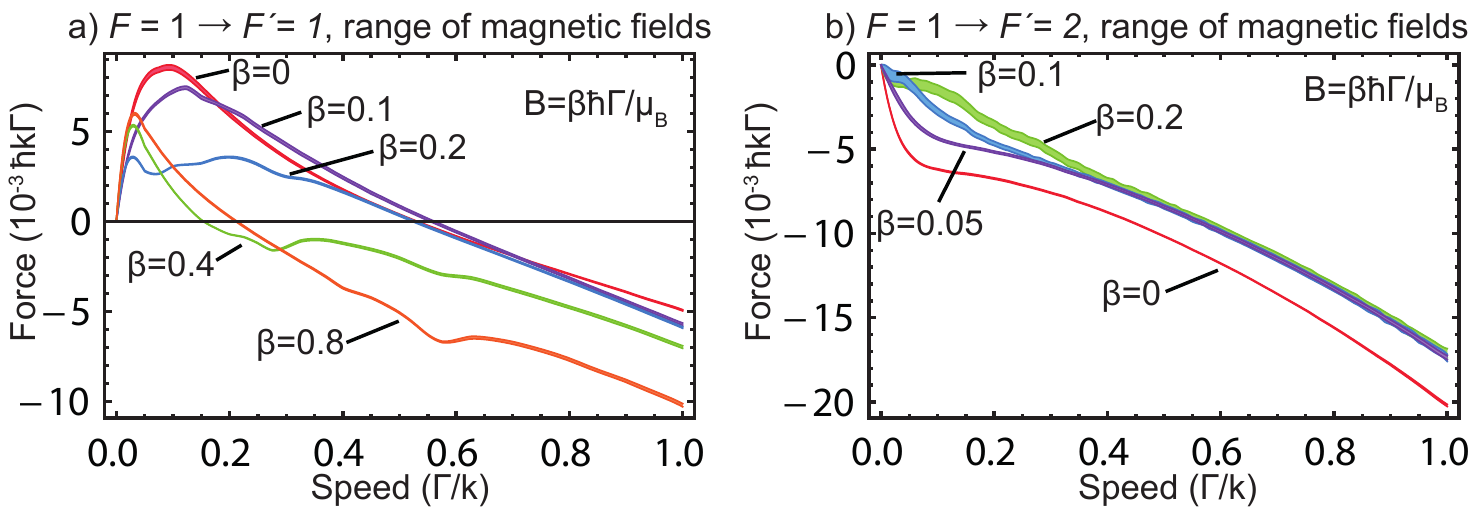}
\end{center}
\caption{Steady-state force in the direction of velocity for a) a $1\rightarrow 1$ system and b) a $1\rightarrow 2$ system moving through a 3D $\sigma^+\sigma^-$ light field with a magnetic field $B=\beta\hbar\Gamma/\mu_B$ applied along $z$. Results are shown for several values of $\beta$. Parameters are $I=I_\textrm{sat}$, $\Delta=-2.5\Gamma$, $g=g'=1$.}
\label{fig:MagFieldResults3D}
\end{figure}

Figure~\ref{fig:MagFieldResults3D}~a) shows the force versus speed in the $1\rightarrow1$ system for a range of magnetic field magnitudes. Notice that both the maximum value of the the sub-Doppler force, and the range of speeds over which it operates, is reduced as the magnetic field strength is increased. Nevertheless, there remains a significant positive force at low velocities for field values up to the maximum explored, $0.8\hbar \Gamma/\mu_B$ corresponding to 0.57\,mT when $\Gamma = 2\pi\times 10$\,MHz. This is very different to the behaviour in the more familiar type-I $1\rightarrow2$ system, shown in figure~\ref{fig:MagFieldResults3D}~b), where the sub-Doppler features present at zero field almost vanish once the field reaches $0.1\hbar \Gamma/\mu_B$. Also visible in figure~\ref{fig:MagFieldResults3D}~a) are a set of weak magnetic field resonances which move up to higher velocities as the magnetic field is increased. For example, when the field is $0.4\hbar \Gamma/\mu_B$ there are resonances in the force when the speed is approximately $0.28\Gamma/k$ and $0.48\Gamma/k$. Exploring the dependence on the g-factors, we find that the force depends on the product of the lower state g-factor, $g$, and the applied field $B$, but is independent of the upper state g-factor, $g'$. For the $1\rightarrow0$ and $2\rightarrow1$ systems the dependence on magnetic field is similar to that of the $1\rightarrow1$ system.

These results lead us to conclude that sub-Doppler processes will be important in typical type-II MOTs, where the atoms typically explore fields up to about $B=0.25 \hbar \Gamma/\mu_B$. Because MOTs use red-detuned light, these sub-Doppler forces drive the velocity upwards and so limit the temperature that can be reached in the MOT. If the intensity is approximately $I_{\rm sat}$, which is a popular choice,  the velocity where the force vanishes corresponds to a temperature far higher than the Doppler limit. This is consistent with experimental observations of high temperatures in type-II MOTs, e.g.~\cite{Raab1987, Norrgard2016}.

\section{Trapping forces}
\label{Trapping}

So far we have investigated the velocity-dependent force. Now we turn to magneto-optical trapping forces for the various angular momentum cases. It is useful to compare our results with those of~\cite{Tarbutt2015}, where the trapping forces in a 3D MOT made of three pairs of $\sigma^+\sigma^-$ beams are calculated for a variety of angular momentum cases using a set of rate equations. Some conclusions of that paper are as follows: (i) when $F'=F$, the polarization handedness required to give a trapping force is the same as for a type-I system, but when $F'=F-1$ the polarization has to be reversed; (ii) the required handedness depends on the sign of the upper state $g$-factor, $g'$, but not on the lower state $g$-factor, $g$; (iii) if $g'=0$ there is no trapping force, and if $g'\ll g$ the trapping force is very weak. In the regime where $|g\mu_BB|>0.5\hbar\Gamma$, the OBE simulations presented below agree with these findings and are in quantitative agreement with the rate model for the few simple cases we have examined. However, in the opposite regime where $|g\mu_BB|<0.5\hbar\Gamma$ they show new behaviour.

To investigate the trapping forces, we again consider the situation where an atom moves with velocity $\boldsymbol{v}$ through a light field $\boldsymbol{\mathcal{E}}_{\sigma^+\sigma^-3\text{D}}$ and a magnetic field $B_q=B\epsilon_0$. We evolve the atomic operators using the OBEs until the solutions reach a quasi-steady state and then calculate the time-average $z$-component of the force $\text{f}_z=\textbf{f}\cdot\textbf{e}_z$. For each choice of magnetic field we fix the speed $|\boldsymbol{v}|$ and average the results over many directions and initial positions. This averaging removes the velocity-dependent part of the force but retains the magneto-optical force. From these results we can infer the position-dependent force in a MOT. For clarity, we plot the force as a function of magnetic field in units of $\hbar \Gamma/\mu_B$, but the reader can convert this into a trapping force once the field gradient is known.
\begin{figure}[tb]
\begin{center}
\includegraphics[scale=1]{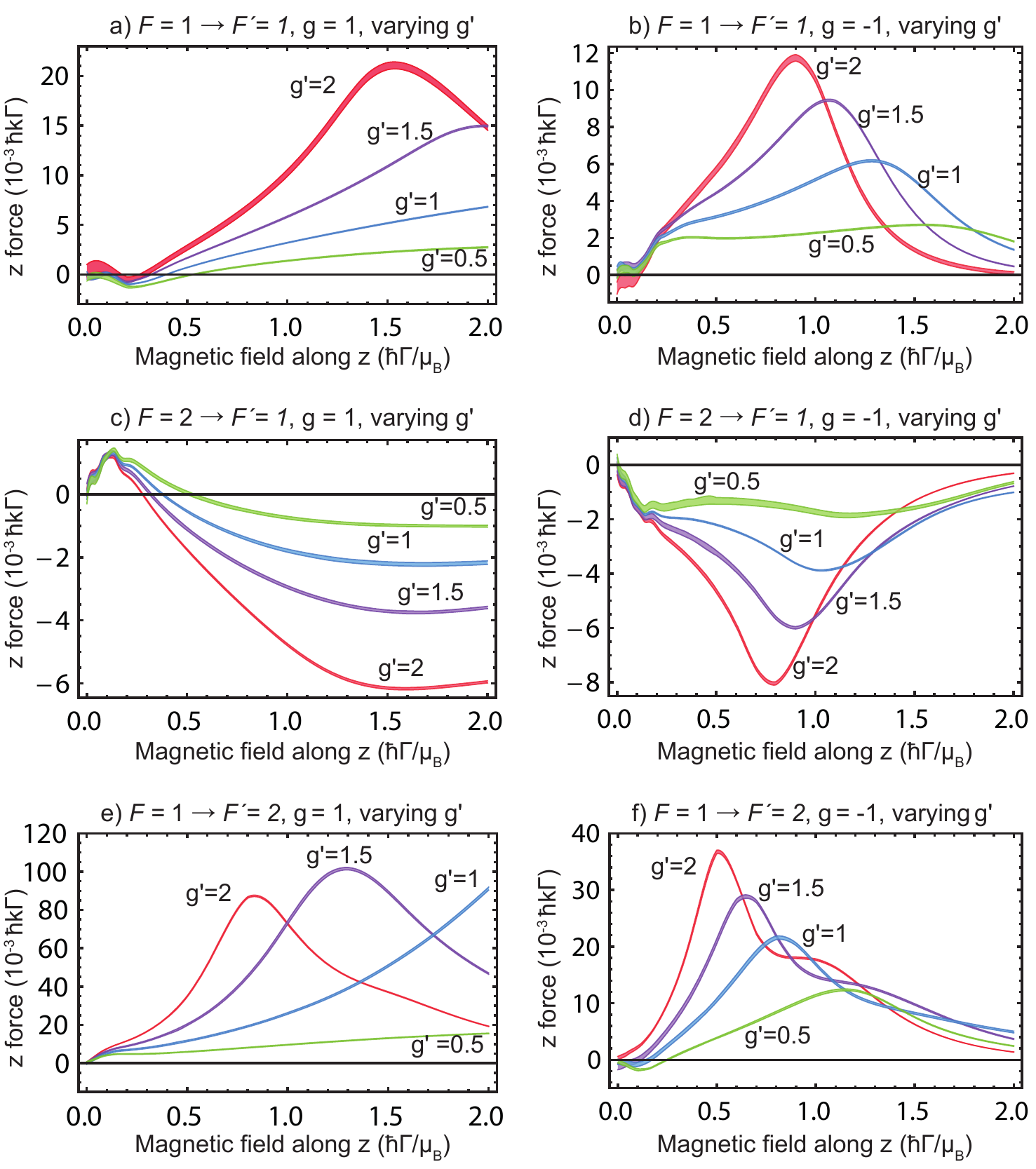}
\end{center}
\caption{Force $\text{f}_z$ versus $B$ (along $z$) for the $1\rightarrow 1$, $2\rightarrow 1$ and $1\rightarrow 2$ systems, for various values of $g'$ and for $g=+1$ (left column) and $g=-1$ (right column). Parameters are $I=I_\textrm{sat}$, $\Delta=-2.5\Gamma$, and $|\boldsymbol{v}|=0.1\Gamma/k$ }
\label{fig:3DTrappingResults}
\end{figure}

The results of the simulations are shown in figure~\ref{fig:3DTrappingResults} where we plot $\text{f}_{z}$ versus $B$ for $g=\pm 1$ and various values of $g'$. At large positive $B$, and for our definition of the light field, equation~(\ref{eq:spsm3dStandingWave}), we see that the force is positive for the $1\rightarrow 1$ and $1 \rightarrow 2$  systems (graphs a), b), e) and f)) but negative for the $2\rightarrow 1$ system (graphs c) and d)). This means that a trapping force requires a negative field gradient for the former systems, but a positive field gradient or reversed polarizations for the latter system. This is the same result as obtained from the rate model~\cite{Tarbutt2015}. For small $B$, we find a change in the behaviour of the force, pointing to a new mechanism. For type-II systems, this new force opposes the normal magneto-optical force when $g$ and $g'$ have the same signs. For type-I systems it is the opposite. In all cases, the ratio of $g'$ to $g$ sets the strength of this new force relative to the usual MOT force. For $|g|>|g'|$   the new force has a greater influence, as can be seen from the green curves in figure~\ref{fig:3DTrappingResults}. This is particularly relevant to molecular MOTs because many molecules of interest have small $g'$. The presence of these new forces at low fields will lead to larger MOT diameters in type-II systems when $g$ and $g'$ have the same sign, especially for the $2\rightarrow 1$ system shown in figure~\ref{fig:3DTrappingResults}~c) where the large anti-trapping force at low fields will force the atoms out to large radii. Repeating these simulations for different speeds $|\boldsymbol{v}|$ in the range $|\boldsymbol{v}|=0$ to $|\boldsymbol{v}|=0.2\Gamma/k$ slightly alters the shape of the trapping features at low magnetic fields but does not affect the observations made above.   

Finally, we note that the trapping force on the $1 \rightarrow 0$ transition, not shown in figure~\ref{fig:3DTrappingResults}, is different from the other type-II transitions - because the upper state has no Zeeman splitting the normal MOT force is absent for this system; we do however find a small force at low magnetic field whose maximum value is about $0.4 \times 10^{-3} \hbar k \Gamma$ for the same choice of parameters as in figure~\ref{fig:3DTrappingResults}.

\section{Conclusions}
\label{MoleculeMOTs}

Let us now summarize our findings, draw some conclusions, and make some suggestions. In 1D, we found that type-II systems moving in a   lin$\phi$lin light field show velocity-dependent forces.  At low velocities blue-detuned light is needed to cool the atoms, whereas at high velocities the light must be red detuned. The force in the low velocity regime is best understood in terms of a cycle of non-adiabatic transitions near the intensity minima, followed by optical pumping from the bright state to the optical dark state near the intensity maxima, which sets up a Sisyphus mechanism. The velocity dependent forces vanish when $\phi=0$ and, in the case of  the $1\rightarrow 1$ and $1 \rightarrow 0$ systems, when  $\phi=\pi/2$. There is no velocity-dependent force when the $1\rightarrow 1$ and $1 \rightarrow 0$ systems traverse a $\sigma^{+}\sigma^{-}$ standing wave, whereas the  $2 \rightarrow 1$ system shows a velocity-dependent force of similar magnitude to the lin$\phi$lin case, but with an opposite sign.

In 3D, all the type-II systems show similar velocity dependent forces in all the light fields considered. At high velocities (``Doppler cooling'' regime), cooling in these systems requires red-detuned light, while at low velocities (``sub-Doppler'' regime) blue-detuned light is needed. We attribute these forces to the non-adiabatic Sisyphus effect, which can occur in both the 3D $\sigma^{+}\sigma^{-}$ and  lin$\perp$lin configurations because all typical trajectories through these light fields involve spatially varying intensities and polarizations. In type-II systems, the sub-Doppler force, and the range of speeds over which it dominates, are reduced when a magnetic field is applied, but to a much smaller extent than in type-I systems. As a result, sub-Doppler processes are likely to play an important role in type-II MOTs as well as in an optical molasses. When the light is red-detuned the Doppler and sub-Doppler forces have opposite signs and so there is a speed, $v_{0}$, where they cancel. Atoms will tend to be driven towards this speed, rather than towards zero. When the intensity in each beam is $I_{\rm{sat}}$, we find $v_{0} \approx 0.5 \Gamma/k$. For Na cooled on the D$_2$ line, this rms speed corresponds to a temperature of about 9\,mK, which is far higher than the Doppler limit. We suggest that this is the reason for the high temperatures observed in type-II MOTs~\cite{Raab1987}, which though observed almost 30 years ago has never been explained. The equilibrium speed $v_{0}$ is proportional to the square root of the laser intensity and so the temperature can be lowered by lowering the intensity once the atoms are captured in the MOT. This is consistent with the recent measurements of temperature versus intensity in a SrF MOT~\cite{Norrgard2016}.

Our study using optical Bloch equations in 3D also reveals a new magneto-optical force at low magnetic field. For type-II systems, this force opposes the normal magneto-optical force when $g$ and $g'$ have the same signs. For this reason, we expect type-II MOTs to be larger than their type-I counterparts when the $g$-factors have the same sign because the trapping forces are reduced, or even reversed, near the trap centre. Investigating the mechanism responsible for this new force is an interesting avenue for future research.

We suggest a three-step approach to obtaining the lowest temperatures for type-II systems. The atoms are first loaded into a normal high-intensity, red-detuned MOT, which will give high capture velocity but also high temperature. The intensity of the lasers can then be ramped down in order to lower the temperature. Then, the detuning is switched to the blue and the intensity ramped back up again so that the sub-Doppler cooling is most effective. This can be done with the magnetic field turned off, following the standard gray molasses method. Alternatively, to retain the trapping forces, the field can be kept on and the polarization handedness reversed when the light is switched from red to blue, making a blue-detuned MOT. We hope the results and insights we have presented will be valuable in the ongoing endeavour to produce high-density samples of ultracold atoms and molecules.

\ack
The research was supported by STFC grant no. ST/N000242 and by EPSRC under grants EP/I012044 and EP/M027716.. Data underlying this article can be accessed through Zenodo (doi: 10.5281/zenodo.168563) and may
be used under the Creative Commons CCZero license. 

\section*{References}

\bibliography{referencesShort}






\end{document}